\documentclass[]{aa}
\usepackage{graphics,latexsym,epsfig,graphicx,txfonts,natbib}
\bibpunct{(}{)}{;}{a}{}{,}

\newcommand{\beqn}{\begin{equation}}
\newcommand{\eeqn}{\end{equation}}

\newcommand{\dd}{{\rm d}}

\begin{document}

\title{Why your model parameter confidences might be too optimistic --\\ unbiased estimation of the inverse covariance matrix}

\author{J.\ Hartlap \and P.\ Simon \and P.\ Schneider}

\institute{
  Argelander-Institut\thanks{Founded by merging of the
  Sternwarte, Radioastronomisches Institut and Institut f\"ur
  Astrophysik und Extraterrestrische Forschung der Universit\"at
  Bonn} f\"ur Astronomie, Universit\"at Bonn, Auf dem H\"ugel 71, D-53121 Bonn, Germany\\
  \email{hartlap@astro.uni-bonn.de}
  }

\date{Received 3 August 2006 / Accepted 24 November 2006}
\authorrunning{Hartlap et al.}
\titlerunning{Unbiased estimation of the inverse covariance matrix}

\abstract{} {The maximum-likelihood method is the standard approach to
obtain model fits to observational data and the corresponding
confidence regions. We investigate possible sources of bias in the
log-likelihood function and its subsequent analysis, focusing on
estimators of the inverse covariance matrix. Furthermore, we study
under which circumstances the estimated covariance matrix is
invertible.}  {We perform Monte-Carlo simulations to investigate the
behaviour of estimators for the inverse covariance matrix, depending
on the number of independent data sets and the number of variables of
the data vectors. } {We find that the inverse of the
maximum-likelihood estimator of the covariance is biased, the amount
of bias depending on the ratio of the number of bins (data vector
variables), $p$, to the number of data sets, $n$. This bias inevitably
leads to an -- in extreme cases catastrophic -- underestimation of the
size of confidence regions. We report on a method to remove this bias
for the idealised case of Gaussian noise and statistically independent
data vectors. Moreover, we demonstrate that marginalisation over
parameters introduces a bias into the marginalised log-likelihood
function. Measures of the sizes of confidence regions suffer from the
same problem.  Furthermore, we give an analytic proof for the fact
that the estimated covariance matrix is singular if $p>n$. } {}

\maketitle

\section{Introduction}
The maximum-likelihood method \citep[e.g.][]{barlow} is common practice to obtain the best-fit parameters $\vec{\pi}_0$ and confidence regions from a measured data vector $\vec{d}\in \mathbb{R}^{p}$ for a model $\vec{m}(\vec{\pi})$. It usually consists of finding the maximum of the log-likelihood function
\begin{equation} \label{logL}
\mathcal{L}(\vec{d}|\vec{\pi})\propto-\frac{1}{2} \left[\vec{d}-\vec{m}(\vec{\pi})\right]^{\mathrm t} \tens{\Sigma}^{-1} \left[\vec{d}-\vec{m}(\vec{\pi})\right]\, ,
\end{equation}
where $\vec{\pi}$ is the parameter vector and a Gaussian distribution of the measurement errors is assumed. The confidence regions for the maximum-likelihood fit are then defined by the surfaces of constant $\Delta \mathcal{L} \equiv \mathcal{L}_{\rm max}-\mathcal{L}$, where $\mathcal{L}_{\rm max}$ is the maximum value of the log-likelihood function. 

For the evaluation of the log-likelihood the population covariance matrix $\tens{\Sigma}$ and its inverse $\tens{\Sigma}^{-1}$ or estimates thereof are needed. In most cases, no exact analytical expression for $\tens{\Sigma}$ can be given, although numerous authors make use of analytical approximations. An example from the field of weak gravitational lensing is \citet{CFHTLSdeep}, who use the Gaussian approximation to the covariance matrix of the shear correlation functions given by \citet{schneiderCov}. Other possibilities are to estimate $\tens{\Sigma}$ from the data themselves \citep[e.g.][]{GABODS6,budavari03} or to obtain it from a simulated data set whose properties are comparable to the original data \citep[e.g.][]{szapudi_mono}. In the latter paper, the authors observed that the estimated covariance matrix becomes singular if $p$, the number of entries of the data vectors, exceeds the number of observations / simulated data vectors. As a remedy, they propose to use the Singular Value Decomposition \citep[SVD, ][]{nr} to obtain a pseudo-inverse of the covariance matrix, but do not investigate the properties of the resulting estimate of $\tens{\Sigma}^{-1}$ in detail.
In this paper, we prove analytically that the rank of the standard estimator of the covariance matrix cannot exceed the number of observations. We then point out that, even if this estimator is not singular, simple matrix inversion yields a biased estimator of $\tens{\Sigma}^{-1}$. This may, if not corrected for, cause a serious underestimate of the size of the confidence regions.	This problem has also been noticed by \citet{HirataGGL} and \citet{MandelbaumIntrinsic}, who use Monte-Carlo simulations to determine the correct confidence contours in cases where the covariance matrix is noisy.
	We report on the existence of a simpler method to remove this bias, which can be derived for Gaussian noise and statistically independent data vectors, and test the validity of this method when these assumptions are violated.

\section{The covariance matrix}
\subsection{Estimators}
Let $\vec{d}$ be a vector of $p$ random variables with components $d_i$, drawn from a multi-variate Gaussian distribution with population covariance matrix $\tens{\Sigma}$ and mean $\vec{\mu}$:
\begin{equation}
P(\vec{d})=\frac{1}{(2\pi)^{p/2}\sqrt{\det \tens{\Sigma}}}\,\exp\left( -\frac{1}{2}(\vec{d}-\vec{\mu})^{\rm t}\tens{\Sigma}^{-1}(\vec{d}-\vec{\mu}) \right)\, .
\end{equation}
Furthermore, let $\vec{d}^{(k)}$ denote the $k$-th realisation of this random vector, where $k\in [1,n]$ and $n$ is the total number of realisations.
The well-known maximum-likelihood estimator for the components of the covariance matrix is given by \citep{barlow}
\begin{equation} \label{Cml}
 \hat{\tens{C}}_{ij}^{\rm ML}=\frac{1}{n}\sum_{k=1}^{n}\, \left(d_i^{(k)}-\mu_i\right) \left(d_j^{(k)}-\mu_j\right) \, ,
\end{equation}
which in the case of a known mean vector $\vec{\mu}$ is unbiased. If, however, \mbox{$\vec{\mu}$} has to be estimated from the data, a correction factor of \mbox{$n/(n-1)$} has to be applied to (\ref{Cml}).

\subsection{The rank of $\hat{\tens{C}}^{\rm ML}$} \label{proof}
In the following, we prove that $\hat{\tens{C}}^{\rm ML}$ is singular for \mbox{$p>n$} in case of known mean vector, and for \mbox{$p>n-1$} if the mean vector is obtained from the data as well.
For the first case, this can be seen by rewriting (\ref{Cml}) as
\begin{equation} \label{Cml2}
 \hat{\tens{C}}^{\rm ML}=\frac{1}{n}\sum_{k=1}^{n}\, \vec{d}^{(k)}\,{\vec{d}^{(k)}}^{\rm t}\, ,
\end{equation}
where we presume, without loss of generality, that the mean vector is zero. Since the data vectors $ \vec{d}^{(k)}$ are statistically independent, we can safely assume that they are linearly independent for \mbox{$n\leq p$} (for a continuous distribution, the probability to draw linearly dependent data vectors is zero). Therefore, $\left\{ \vec{d}^{(k)} \right\}$ span an $n$-dimensional subspace $U$ of $\mathbb{R}^p$. To check whether $\hat{\tens{C}}^{\rm ML}$ is singular we now try to find a vector $\vec{y}\neq \vec{0}$ for which \mbox{$\hat{\tens{C}}^{\rm ML}\,\vec{y}=\vec{0}$}. 
Looking at (\ref{Cml2}), we see that this is only possible for \mbox{$p>n$}, since in this case we can always choose a vector $\vec{y}$ from the subspace orthogonal to $U$, for which \mbox{${\vec{d}^{(k)}} \cdot \vec{y}=0\; \forall\; k$}. 
If $p\leq n$,  $\left\{ \vec{d}^{(k)} \right\}$ already spans the whole of $\mathbb{R}^p$, and no vector can be found that is orthogonal to all $\vec{d}^{(k)}$. This proves that $\hat{\tens{C}}^{\rm ML}$ is singular for known mean vector if $p>n$.\\

We now prove our statement for an unknown mean vector $\vec{\mu}$, which is estimated from the data using
\beqn
 \vec{\mu}=\frac{1}{n}\,\sum_{k=1}^{n}\,	\vec{d}^{(k)}\; .
\eeqn
For this, we define a new set of independent data vectors \mbox{$\left\{ \vec{w}^{(k)} \right\}$} by forming linear combinations of $\left\{ \vec{d}^{(k)} \right\}$, specified by the orthogonal transformation $\tens{B}$, of which we demand that the last ($n$-th) row be given by \mbox{$(1/\sqrt{n},\, \ldots \, , 1/\sqrt{n})$} \citep{TWA}:
\begin{equation} \label{orthotrans}
 \vec{w}^{(k)} = \sum_{l=1}^{n}\, \tens{B}_{kl}\, \vec{d}^{(l)} \; . 
\end{equation}
Thanks to our choice of $\tens{B}_{nl}$, we have \mbox{$\vec{w}^{(n)}=\sqrt{n}\,\vec{\mu}$}.
Next, we rewrite $\hat{\tens{C}}^{\rm ML}$ by means of the new data vectors:
\begin{eqnarray}
 \hat{\tens{C}}^{\rm ML}&=&\frac{1}{n}\sum_{k=1}^{n}\, \vec{d}^{(k)}\,{\vec{d}^{(k)}}^{\rm t} - \vec{\mu}\vec{\mu}^{\rm t}\\
 &=& \frac{1}{n}\sum_{k=1}^{n}\, \vec{w}^{(k)}\,{\vec{w}^{(k)}}^{\rm t} - \frac{1}{n}\vec{w}^{(n)}\,{\vec{w}^{(n)}}^{\rm t}\\
 &=& \label{Ctrafo} \frac{1}{n}\sum_{k=1}^{n-1}\, \vec{w}^{(k)}\,{\vec{w}^{(k)}}^{\rm t}\; .
\end{eqnarray}
The last expression is of the same form as (\ref{Cml2}) (except for the sum, which has one addend less), and so the same line of reasoning as above can be applied to show that $\hat{\tens{C}}^{\rm ML}$ is singular for \mbox{$p>n-1$}.\\

Another interesting implication of Eq.\ (\ref{Ctrafo}) is that the mean vector and the estimated covariance matrix are distributed independently \citep[again see][]{TWA}, although they are computed from the same data vectors. First, note that $\vec{w}^{(i)}$ and $\vec{w}^{(j)}$ are statistically independent for \mbox{$i\neq j$}. This can be seen by computing the covariance between the two vectors:
\begin{eqnarray}
{\rm Cov}\left(\vec{w}^{(i)},\vec{w}^{(j)}\right) &=&  \left\langle \left(\vec{w}^{(i)}-\vec{\nu}^{(i)}\right)\left(\vec{w}^{(j)}-\vec{\nu}^{(j)}\right)^{\rm t} \right\rangle\\
 &=& \sum_{k,l=1}^{n}\, \tens{B}_{ik} \tens{B}_{jl}\, \left\langle \left(\vec{d}^{(k)}-\vec{\mu}\right)\left(\vec{d}^{(l)}-\vec{\mu}\right)^{\rm t} \right\rangle\\
  &=& \sum_{k,l=1}^{n}\,\tens{B}_{ik} \tens{B}_{jl}\,\delta_{kl}\,\tens{\Sigma}\\
 &=& \delta_{ij}\,\tens{\Sigma}
\end{eqnarray}
Here, $\langle \cdot \rangle$ denotes the expectation value and (see
Eq. \ref{orthotrans})
\begin{equation}
  \vec{\nu}^{(i)}=\left\langle\vec{w}^{(i)}\right\rangle= 
  \sum_{j=1}^{n}\, \tens{B}_{ij}\,\left\langle\vec{d}^{(j)}\right\rangle=   
  \vec{\mu} \sum_{j=1}^{n}\tens{B}_{ij}
\end{equation}
is the mean value of $\vec{w}^{(i)}$.

Since $\hat{\tens{C}}^{\rm ML}$ does not depend on $\vec{w}^{(n)}$, which in turn is statistically independent of the remaining $\vec{w}^{(i)}$, this shows the independence of estimated mean and covariance.

\section{The inverse covariance matrix}
\subsection{An unbiased estimator for $\tens{\Sigma}^{-1}$}

From (\ref{Cml}), an estimator for $\tens{\Sigma}^{-1}$ can be obtained by matrix inversion:
\begin{equation}  \label{biased}
 \hat{\tens{C}}^{-1}_*= \left(\hat{\tens{C}}^{\rm ML}\right)^{-1}\, .
\end{equation}
This estimator is consistent, but \emph{not unbiased} due to noise in $\hat{\tens{C}}^{\rm ML}$: the inverse of an unbiased estimator for some statistical variable $X$ is in general not an unbiased estimator for $X^{-1}$. Indeed, in our case of Gaussian errors and statistically independent data vectors one can show \citep{TWA} that the expectation value of $\hat{\tens{C}}^{-1}_*$ is \emph{not} the inverse of the population covariance, but
\begin{equation} \label{invCexpectation}
 \left\langle \hat{\tens{C}}^{-1}_* \right\rangle = \frac{N}{N-p-1}\; \tens{\Sigma}^{-1}\; \mbox{for}\; p<N-1\, ,
\end{equation}
where $N=n$ if $\vec{\mu}$ is known and \mbox{$N=n-1$} if the mean is estimated from the data. In the following, we will only pursue the latter case.\\

The amount of bias in $\hat{\tens{C}}^{-1}_*$ thus depends essentially on the ratio of the number of entries $p$ in the data vectors (henceforth referred to as the number of bins) to the number of independent observations $n$.
It leads to an underestimation of the size of confidence regions by making the log-likelihood function steeper, but it does not change the maximum-likelihood point and thus does not affect the parameter estimates themselves.

From (\ref{invCexpectation}) it follows that an unbiased estimator of $\tens{\Sigma}^{-1}$ is given by\footnotemark[1]
\begin{equation} \label{invCunbiased}
 \hat{\tens{C}}^{-1} = \frac{n-p-2}{n-1}\; \hat{\tens{C}}^{-1}_* \; \mbox{for}\; p<n-2\, .
\end{equation}

\begin{figure}[t]
\resizebox{\hsize}{!}{\includegraphics[angle=0]{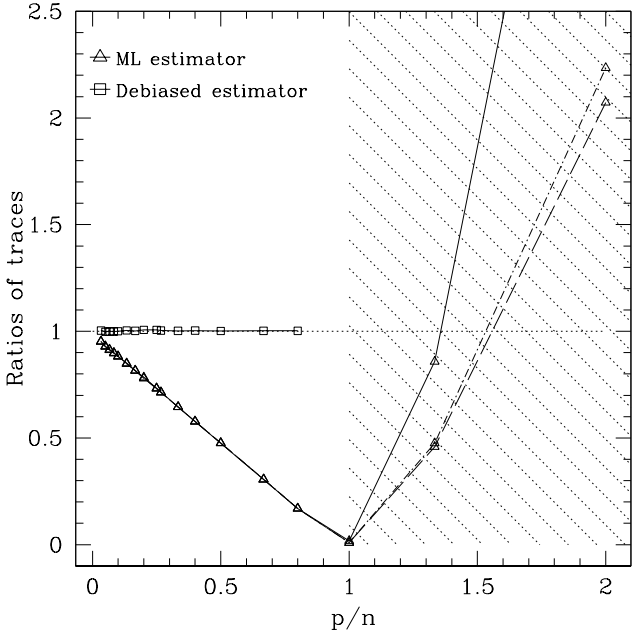}}
\caption{Ratios of the trace of $\tens{\Sigma}^{-1}$ to the traces of $\tens{C}^{-1}_*$ (triangles) and $\hat{\tens{C}}^{-1}$ (squares), respectively. The dashed line is for the covariance model (\ref{Sigdc}), the solid line for (\ref{Sigdl}) and the dot-dashed-line for (\ref{Signd}). The original data vectors had \mbox{$p_1=240$} bins, and were rebinned by subsequently joining $2,\, 3,\, \ldots$ of the original bins. The number of independent observations is \mbox{$n=60$}. Error bars are comparable to the symbol size and therefore omitted.}
\label{covbias}
\end{figure}

\footnotetext[1]{
Note that there is a typing error in Anderson's book, where he gives an expression corresponding to
\[
 \hat{\tens{C}}^{-1} = \frac{n-p-2}{n-\vec{2}}\; \hat{\tens{C}}^{-1}_* \; \mbox{for}\; p<n-2.
\]
}

\subsection{Monte-Carlo experiments}

To illustrate Eq.\ (\ref{invCunbiased}), and also to probe how the pseudo-inverse of the estimated covariance obtained by the Singular Value Decomposition behaves (see below), we perform the following experiment:
First, we choose an analytical form for the population covariance $\tens{\Sigma}$. We use three different models:
\begin{eqnarray}
	\label{Sigdc} \tens{\Sigma}^{\rm d,c}_{ij} &=& \sigma^2\,\delta_{ij}\;,\\ 
	\label{Sigdl} \tens{\Sigma}^{\rm d,l}_{ij} &=& \sigma^2\,\left[1-i/(1+p_1)\right]\,\delta_{ij}\;\;\;\mbox{and}\\
	\label{Signd} \tens{\Sigma}^{\rm nd}_{ij}  &=& \sigma^2/(1+\epsilon |i-j|) \;\; ,
\end{eqnarray}
which initially are \mbox{$p_1\times p_1$} matrices. $\epsilon$ can be used to tune the degree of correlation in model (\ref{Signd}); we choose \mbox{$\epsilon=0.05$}. 

We then create $n$ data vectors of length $p_1$ according to \mbox{$\vec{d}^{(k)} = \vec{m} + \vec{\gamma}^{(k)}\left(\tens{\Sigma}_1\right)$}, where \mbox{$\vec{\gamma}^{(k)}\left(\tens{\Sigma}_1\right)$} is a noise vector drawn from a multivariate Gaussian distribution with mean zero and covariance $\tens{\Sigma}_1$. The choice of the model vector $\vec{m}$ is arbitrary, and in fact for the present purpose it would be sufficient to set \mbox{$\vec{m}=\vec{0}$}. For later use, however, we choose the linear model \mbox{$m_i=ax_i + b$}, where \mbox{$x_i=(x_{\rm max}-x_{\rm min})(i+1/2)/p_1$} is the value of the free variable corresponding to the centre of the $i$-th bin.

From this synthetic set of observations we estimate the mean data vector and the covariance matrix, which yields the estimator $\hat{\tens{C}}^{\rm ML}$. Next, both $\tens{\Sigma}$ and $\hat{\tens{C}}^{\rm ML}$ are inverted using the Singular Value Decomposition (see below). Finally, we compute the unbiased estimate $\hat{\tens{C}}^{-1}$ of the inverse covariance as given in (\ref{invCunbiased}).

To probe the dependence of the bias of the estimators for
$\tens{\Sigma}^{-1}$, $n$ new data vectors are created subsequently
with \mbox{$p_j=p_1/j$} bins, for all integer \mbox{$j\in[2,\ldots,p_1/2]$}, where the population covariance $\tens{\Sigma}_j$ for $p_j$ bins can be obtained from the original $\tens{\Sigma}_1$ by averaging over \mbox{$(j\times j$)}-sub-blocks of $\tens{\Sigma}_1$.
This strategy of re-binning has the advantage that the true covariance is known exactly for all $p_j$.

Since the bias in Eq.\ (\ref{invCexpectation}) is just a scalar factor, we record the traces of the estimators $\hat{\tens{C}}^{-1}_*$ and $\hat{\tens{C}}^{-1}$ for each number of bins $p$. To improve our statistics, we repeat the procedure outlined above $10^4$ times and average over the traces computed in each step.\\

In Fig.\ \ref{covbias}, we plot the ratios of the trace of $\tens{\Sigma}^{-1}$ to the traces of $\hat{\tens{C}}^{-1}_*$ and $\hat{\tens{C}}^{-1}$, respectively. Not using the bias-corrected $\hat{\tens{C}}^{-1}$ can have considerable impact on the size of confidence regions of parameter estimates: for \mbox{$p<n-2$}, the components of $\hat{\tens{C}}^{-1}_*$ will be too large compared to those of the true inverse covariance, and the log-likelihood will decrease too steeply, resulting in confidence contours too small.

 We also plot the traces of $\hat{\tens{C}}^{-1}_*$ for the different covariance models beyond \mbox{$p\geq n-1$}, where the estimator $\hat{\tens{C}}^{\rm ML}$ is singular. These data points have been obtained using the Singular Value Decomposition to invert the covariance matrix, yielding a decomposition of the form
\begin{equation}
 \tens{C} = \tens{U} \tens{W} \tens{V}^{\rm t}\; ,
\end{equation}
where $\tens{U}$ and $\tens{V}$ are orthogonal matrices and $\tens{W}$ is a diagonal matrix containing the singular values. Since $\tens{C}$ is symmetric, one has in addition \mbox{$\tens{U} = \tens{V}$}, while $\tens{W}$ contains the moduli of the eigenvalues of $\tens{C}$.
The inverse of $\tens{C}$ is then given by \mbox{$\tens{C}^{-1}= \tens{V} \tens{W}^{-1} \tens{U}^{\rm t}$}. If $\tens{C}$ is singular,
some of the entries of $\tens{W}$ will be zero or comparable to machine precision. We therefore can only compute a pseudo-inverse of $\tens{C}$ by replacing the inverses of these singular values in $\tens{W}^{-1}$ by zero, as has been suggested in \citet{nr} and \citet{szapudi_mono}. Fig.\ \ref{covbias} shows that the bias of $\hat{\tens{C}}^{-1}_*$ in this regime depends significantly on the covariance model chosen and does not depend on binning in a simple way. Therefore we strongly discourage from the use of the SVD for \mbox{$p>n-1$}.

\section{Implications for likelihood analysis}

Having obtained an unbiased estimator of the inverse covariance
matrix, one may still be concerned about a possible bias in the
log-likelihood function, since it consists of the product of
\mbox{$\left(\vec{d}-\vec{\mu}\right)$} and $\hat{\tens{C}}^{-1}$
(Eq.\ \ref{logL}), since $\vec{\mu}$ and $\hat{\tens{C}}^{-1}$ are estimated from the same set of observations. In other words, the question is if it is possible to write
\begin{eqnarray} \label{lexp}
 \left\langle \mathcal{L}(\vec{d}|\vec{\pi}) \right\rangle &=& -\frac{1}{2}\;\left\langle  \left(\vec{\mu}-\vec{m}\right)^{\rm t} \hat{\tens{C}}^{-1} \left(\vec{\mu}-\vec{m}\right) \right\rangle\\ 
 &=& -\frac{1}{2}\; \left(\left\langle\vec{\mu}\right\rangle-\vec{m}\right)^{\rm t} \left\langle \hat{\tens{C}}^{-1}\right\rangle \left(\left\langle\vec{\mu}\right\rangle-\vec{m}\right)^{} \;.
\end{eqnarray}
Luckily, this is indeed the case, since we have shown at the end of Sect.\ \ref{proof} that mean vector and covariance matrix are distributed independently.\\

\subsection{Marginalised likelihood}
Usually, one is not only interested in the full parameter space of a problem, but also in values of single parameters and the corresponding errors. In the Bayesian framework, this is usually achieved by marginalising over the ``uninteresting'' parameters: the log-likelihood function $\mathcal{L}_i$ for the single parameter $\pi_i$ is computed using
\begin{equation}
 \mathcal{L}_i(\vec{d}|\pi_i) =\log\left\{ \left[ \prod_{j\neq i} \int\dd \pi_j\right] \; \exp\left[ \mathcal{L(\vec{d}|\vec{\pi})}\right]\right\}  \;.
\end{equation}
There is no reason to believe that the marginalised log-likelihood,
which is a highly non-linear function of the (unbiased) estimate of
the full log-likelihood, and with it the size of the errors on $\pi_i$
are unbiased, even if one uses the unbiased $\hat{\tens{C}}^{-1}$. We
demonstrate this by means of our simulated fitting procedure, where we
now use in addition to the straight line model also a second
simulation using a power-law model of the form
\mbox{$m_i=a\,x_i^b$}. We marginalise over the intercept of the line
and the power-law index, respectively.
We record the sums over all pixels
of the (one-dimensional) grid of the marginalised log-likelihood
functions, which we compute using the true $\tens{\Sigma}^{-1}$ and
the unbiased estimator $\hat{\tens{C}}^{-1}$.
For $\tens{\Sigma}$, we choose the model (\ref{Sigdc}).
We average over \mbox{$\approx 3\times 10^4$}
repetitions of these experiments.  We plot the ratio of true to
estimated log-likelihood sums in Fig.\ \ref{like2fig} (triangles and
solid lines). The plot shows a bias of maximally \mbox{$\approx 8\%$}
for the straight line and even less for the power-law, in a direction
which would lead to an overestimation of the error bars on slope and
amplitude. Although the effect is not very large, this is not
guaranteed to remain so for models different from the ones considered
here.

\subsection{Measuring the size of confidence regions}
For some applications, it is useful to have a simple measure of the size of the confidence regions. As an example, we make use of the Fisher information matrix $\tens{F}$ \citep{fisher}, which is defined by \citep{TTH}
\beqn
 \tens{F}\equiv \left\langle  \frac{\partial^2 \mathcal{L}}{\partial \pi_i\;\partial \pi_j}\right\rangle \; ,
\eeqn
where the derivative is to be evaluated at the maximum-likelihood point $\vec{\pi}_0$.
$\tens{F}$ can be interpreted as an estimate of the inverse covariance matrix of the parameter estimates, provided $\mathcal{L}$ is well approximated by a Gaussian around the maximum-likelihood point.

To demonstrate the bias in \mbox{$\sqrt{\det \tens{F}^{-1}}$}, we compute the Fisher matrix for the straight line and power-law fits using \citep{TTH}
\beqn
 F_{ij}=\sum_{\alpha,\beta=1}^p\; \frac{\partial m_\alpha}{\partial \pi_i} \frac{\partial m_\beta}{\partial \pi_j}\;\tens{C}^{-1}_{\alpha\beta}\; ,
\eeqn
which is valid if the covariance matrix does not depend on the parameters $\pi_i$; $\vec{m}$ is the model vector.

\begin{figure}[t]
\resizebox{\hsize}{!}{\includegraphics[angle=0]{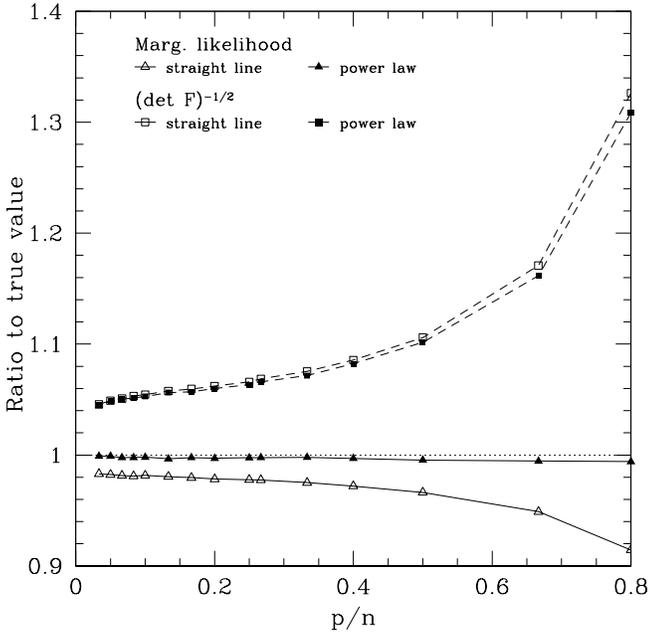}}
\caption{\emph{Triangles, solid lines:} Ratio of the sum over all
pixels of the marginalised likelihood computed using
$\hat{\tens{C}}^{-1}$ and the true marginalised likelihood. Filled
triangles are for the power-law fit (marginalised over the power-law
index), open triangles are for the straight line fit (marginalised
over the intercept). \emph{Squares, dashed lines:} Ratio of
$\sqrt{\det\tens{F}^{-1}}$ using $\hat{\tens{C}}^{-1}$ to the true
one, computed with $\tens{\Sigma}$. For both cases
$\tens{\Sigma}=\tens{\Sigma}^{\rm d,c}$.}
\label{like2fig}
\end{figure}

In Fig.\ \ref{like2fig}, we give the ratio of $\sqrt{\det \tens{F}^{-1}}$, computed using the unbiased estimated covariance $\hat{\tens{C}}^{-1}$, to the value computed using the true covariance (boxes and dashed lines). One sees that in this case the size of the confidence regions is significantly overestimated, for $p/n$ approaching unity by as much as $\approx 30 \%$ for the straight line case, and by a comparable, albeit slightly smaller factor for the power-law fit.

\section{Bootstrapping and non-Gaussian statistics}

\begin{figure}[t]
\resizebox{\hsize}{!}{\includegraphics[angle=0]{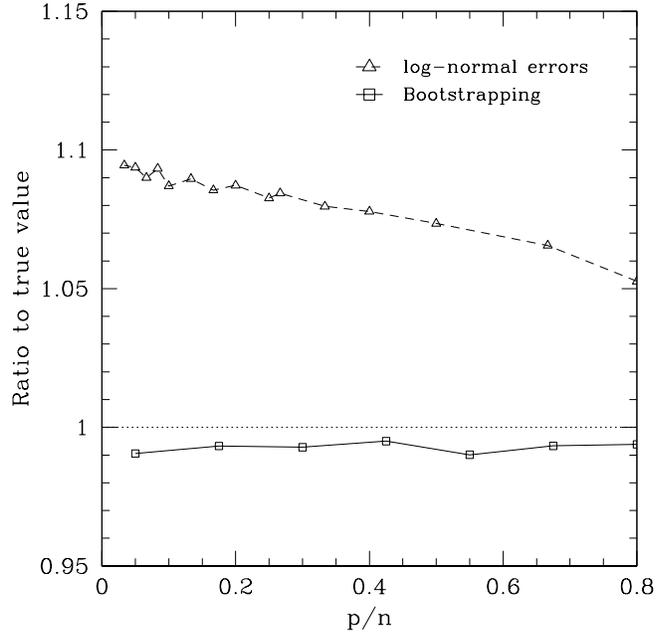}}
\caption{Ratio of the traces of the unbiased estimator $\hat{\tens{C}}^{-1}_*$ and $\hat{\tens{C}}^{-1}$ to the trace of $\tens{\Sigma}^{-1}$; the covariances for the solid curve have been estimated using bootstrapping (see text), the dashed line shows the ratio of the traces for log-normal errors.}
\label{bsngfig}
\end{figure}

The derivation of the unbiased estimator $\hat{\tens{C}}^{-1}$ rests on the assumptions of Gaussian noise and statistically independent data vectors. To test the performance of this estimator in real world situations, where one or both of these assumptions may be violated, we make use of an example from the domain of weak gravitational lensing. For an introduction to this field we refer the reader to \citet{bs}.

We simulate a weak lensing survey consisting of one single field, containing $N_{\rm g}$ galaxies, which are assigned a random ellipticity $\epsilon$. \mbox{$\epsilon=\epsilon_1+{\rm i}\epsilon_2$} is a complex number, which is related to the quadrupole moment of the light distribution of a galaxy \citep[see][]{bs}.
The two components of the ellipticity are drawn from a Gaussian distribution with dispersion $\sigma_\epsilon/\sqrt{2}$.

The goal of the survey is to measure the shear correlation function $\xi_+(\vartheta)$ and to fit a model prediction to it. An estimator for $\xi_+$ is given by \citep{schneiderCov}
\begin{equation}
 \hat{\xi}_+(\vartheta)=\frac{\sum_{ij}\;\left(\epsilon_{1}^{(i)}\epsilon_{1}^{(j)}+\epsilon_{2}^{(i)}\epsilon_{2}^{(j)}\right)\;\Delta_\vartheta\left( |\vec{\theta}_i-\vec{\theta}_j|\right)}{2n_{\rm p}(\vartheta)} \; ,
\end{equation}
where the galaxies are labelled with $i$ and $j$ and have the angular positions $\vec{\theta}_i$ and $\vec{\theta}_j$. $\Delta_\vartheta(\phi)$ is unity if $\vartheta-\Delta\vartheta/2 < \phi \leq \vartheta+\Delta\vartheta/2$, where $\Delta\vartheta$ is the bin width, and zero otherwise. Finally, $n_{\rm p}(\vartheta)$ is the number of pairs of galaxies contributing to the correlation function in the bin centred on $\vartheta$.

We also need the covariance matrix of $\xi_+(\vartheta)$, which, since we only have one measurement, is estimated using the bootstrapping algorithm \citep[e.g.][]{efronBS}: First, we create a catalogue of all \mbox{$N_{\rm p}=N_{\rm g}(N_{\rm g}-1)/2$} possible pairs of galaxies in the field. We then create $N_{\rm bs}$ bootstrap realisations of the survey by repeatedly drawing $N_{\rm p}$ pairs with replacement from the catalogue. From these, we estimate the mean data vector and the covariance matrix of the shear correlation function. As before, we do this for various numbers of bins, where we record the dependence of the traces of $\tens{\Sigma}^{-1}$, $\hat{\tens{C}}^{-1}_*$ and $\hat{\tens{C}}^{-1}$ on binning. For the simple case of pure shape noise, the population covariance is diagonal and can be easily computed using \citep{schneiderCov}
\beqn
 \tens{\Sigma}_{ij} = \frac{\sigma_\epsilon^4}{2\, n_{\rm p}(\vartheta_i)} \; \delta_{ij}\;\; ,
\eeqn
where $\vartheta_i$ is the angular separation corresponding to the centre of the $i$-th bin.
We precompute the function $n_{\rm p}$ numerically from a large set of independent data fields for all binning parameters we wish to use in the simulation.

In principle, both of the assumptions made for the derivation of Eq.\ (\ref{invCunbiased}) are violated: The noise in the shear correlation function is $\chi^2$-distributed, because \mbox{$\xi_+ \propto \epsilon\epsilon$}, where $\epsilon$ is drawn from a Gaussian. However, the number of degrees of freedom of the $\chi^2$-distribution, which equals the number of pairs, is very large, so that it is very well approximated by a Gaussian (central limit theorem). We therefore do not expect any significant influence on the performance of $\hat{\tens{C}}^{-1}$. We expect a larger impact by the fact that the data vectors resulting from the bootstrapping procedure are not statistically independent, since different bins necessarily contain the identical galaxy pairs. Strictly speaking, also the requirements for the application of the bootstrap procedure are not met, since the pairs of galaxies which we use to sample the distribution of the shear correlation function are not statistically independent. However, we argue that drawing individual galaxies instead of pairs is not correct, since this would sample the distribution of $\epsilon$, not the one of $\xi_+$.

The outcome of $\approx 2\times 10^4$ realisations of this experiment is given in Fig.\ \ref{bsngfig} (solid line), with \mbox{$N_{\rm g}=500$} and \mbox{$N_{\rm bs}=40$}. The figure shows that, in spite of the correlations among the pairs of galaxies and the data vectors, $\hat{\tens{C}}^{-1}$ is wrong by only \mbox{$\approx 1\%$}, and may well be used in bootstrap applications like this. 

Finally, we explore the impact of non-Gaussian noise. For this
purpose, we perform the same experiment as before, only replacing the
Gaussian noise vectors ones, $\gamma^{(k)}$, with a log-normal
distribution. These are computed using \beqn
\gamma^{(k)}_i=\exp\left(r^{(k)}_i -1/2\right)\; , \eeqn where
$\vec{r}^{(k)}$ are vectors containing uncorrelated, Gaussian random
variables with mean zero and variance \mbox{$\sigma^2=1$}.  The result
is shown in Fig.\ \ref{bsngfig} (dashed line). Clearly, Eq.\ (\ref{invCunbiased}) is
no longer applicable, although for \mbox{$p/n>0.2$}, one still does
much better with it than without it.

\section{Summary and conclusions}
We have given a proof for the fact that the standard estimator of the covariance matrix (\ref{Cml}) is singular for \mbox{$p>n$} if the mean vector used in (\ref{Cml}) is known, and for \mbox{$p>n-1$} if the mean is estimated from the same data set as the covariance matrix. 
Furthermore, we noted that the inverse of the maximum-likelihood
estimator of the covariance matrix is a biased estimator of the
inverse population covariance matrix. This bias depends basically on
the ratio of the number of bins to the number of independent
observations and can be quite severe as these two numbers become
comparable. If uncorrected for, it will lead to a significant
underestimation of the size of confidence regions derived from
maximum-likelihood fits. The bias can be corrected for \mbox{$p<n-2$}
by using the estimator (\ref{invCunbiased}) instead, which was derived by \citet{TWA} under the assumption of Gaussian errors and statistically independent data vectors.
We stress that there is no contradiction between the foregoing two statements: The singularity of $\hat{\tens{C}}_{\rm ML}$ for \mbox{$p>n-1$} derives from linear algebra alone, whereas the fact that $\hat{\tens{C}}^{-1}$ is zero for \mbox{$p=n-2$} is due to the statistical distribution of the covariance matrix. 
Going beyond \mbox{$p=n-1$}, we find that it is not advisable to use the Singular Value Decomposition to invert the estimated covariance matrix, since the bias of the pseudo-inverse does not seem to be controllable and depends strongly on the population covariance matrix, which is a-priori unknown.

Given the unbiased estimator $\hat{\tens{C}}^{-1}$, we argue that also the log-likelihood function is unbiased. However,  great care has to be taken if one wishes to perform further analysis of $\mathcal{L}$: since it is a statistical variable, any nonlinear operation on it has the potential to cause a bias. We demonstrate this for the case of marginalisation over certain parameters, where the bias is relatively mild for the examples we chose. The situation is much worse if one tries to quantify the size of the confidence regions. The square root of the determinant of the inverse Fisher matrix shows a significant amount of bias, and therefore should not be used as absolute measures of measurement uncertainty.
  
The upshot of all this is the following: avoid to use more bins for your likelihood fit than you have realisations of your data. If your errors are Gaussian and the data vectors are statistically independent, use the estimator $\hat{\tens{C}}^{-1}$ to obtain an unbiased estimate of the inverse covariance matrix and the log-likelihood function. If one or both of these two requirements are not fulfilled, the estimator is not guaranteed to work satisfactorily; this should be checked from case to case. 
  
Finally, we note that the estimates of the covariance matrix and its inverse can be quite noisy. If one has prior knowledge about the structure of the covariance matrix, one can develop estimators with a much lower noise level. Since this noise is responsible for most of the problems discussed in this paper, these improved estimators may also be useful in situations where the requirements for the use of $\hat{\tens{C}}^{-1}$ are not fulfilled. We will explore these possibilities in a future paper.

\begin{acknowledgements}
We wish to thank Tim Eifler for useful suggestions, Jack Tomsky for
pointing us to Anderson's book and Jens R\"odiger, Martin Kilbinger,
Ismael Tereno and Marco Hetterscheidt for careful reading of the manuscript.
\end{acknowledgements}


\bibliographystyle{aa}
\bibliography{6170}

\end{document}